\documentclass{article}
\usepackage{spconf,amsmath,graphicx,hyperref}
\usepackage{amssymb}
\title{SLAP: Learning Speaker and Health-Related Representations from Natural Language Supervision}
\name{Angelika Ando, Auguste Crabeil, Adrien Lesage, Rachid Riad}
\address{Callyope, Paris, France}

\begin{document}
%\ninept
%
\maketitle
\begin{abstract}

Speech encodes paralinguistic information such as demographics, voice quality, and health. Yet no audio foundation model supports zero-shot or out-of-distribution (OOD) generalization to these tasks. We introduce SLAP (Speaker contrastive Language-Audio Pretraining), the first model aligning speech with natural language descriptions of speaker and health metadata through contrastive learning. SLAP combines a Vision Transformer audio encoder with text encoders, trained on more than 3400 hours across 9 datasets with diverse speaker annotations. We evaluated on 38 binary classification tasks spanning demographics, voice characteristics, and clinical assessments across 14 datasets in 7 languages. SLAP achieves 62.9\% average F1 in zero-shot evaluation, a 48\% relative improvement over CLAP (42.4\%), while demonstrating strong OOD generalization to unseen languages and clinical populations.  When fine-tuned with linear probing, SLAP reaches 69.3\% F1 overall and achieves best-in-class performance on health tasks (57.9\% F1), surpassing larger foundation models. 

\end{abstract}

\vspace{-0.5em}
\section{Introduction}
\vspace{-0.5em}
\label{sec:intro}

Speech conveys rich paralinguistic information beyond lexical content, revealing who is speaking and how they sound through acoustic cues. \cite{narayanan2013behavioral}. These cues encode an important number of speaker attributes including age, sex, emotional state, language background most importantly health status, making speech a non-invasive modality into human physiology, mental health and cognition \cite{cummins2018speech}. The capability to extract these attributes at scale could greatly help the healthcare delivery system, enabling easy remote monitoring, early detection of relapses \cite{dorsey2018teleneurology}, and personalized interventions across diverse vulnerable populations and languages. 

While foundation models like Whisper \cite{radford2023robust} allowed great progress for speech recognition and semantic understanding,  they did not integrate any mechanisms to leverage speaker metadata for representation learning or improved inference \cite{wang2024overview}. Although recent models such as SALMONN \cite{tang2024salmonn} have made progress toward modeling certain speaker characteristics, current audio foundation models remain limited when evaluated on speaker-centric tasks \cite{wang2024overview}. For instance, SALMONN freezes its audio encoders, which may constrain the richness of speaker representations and limit their usefulness for tasks such as age, sex, or language recognition, as well as health-related applications. 

Current approaches to speech representation learning face fundamental limitations for clinical deployment and wider distribution. Self-supervised methods such as Wav2Vec2 \cite{baevski2020wav2vec} or MAE-style pretraining \cite{he2022masked} learn powerful acoustic representations by predicting masked segments or reconstructing spectrograms. Yet, without any finetuning data they offer little capacity to infer attributes such as age, sex, language, or health-related indicators, especially in low-data regimes.

Some of these limitations have been overcome in the vision community with the introduction of the CLIP \cite{radford2021learning}, by showing that aligning images and text descriptions could enable zero-shot inference on novel questions. This paradigm has been adapted to audio through CLAP \cite{elizalde2023clap, wu2023large}, where training aligns audio and descriptions but uses front-ends unsuited to paralinguistic cues.

Recent variants such as ParaCLAP \cite{jing24b_interspeech} and EmotionRankCLAP  \cite{chandra25_interspeech} have begun to adapt CLAP for paralinguistic tasks, mainly targeting emotion recognition, and \cite{shih2023speechclip} developed speechCLIP for scene retrieval. While these studies illustrate the potential to combine more adapted speech encoders with language supervision, they remained limited to emotional state and do not address the broader range of speaker and health attributes. 

These diverse efforts show the potential of language supervision for audio, but they have so far been explored only in some specific contexts such as recognition, verification, or emotion. This leaves open the opportunity to build a more general foundation model  that captures demographics, voice quality, and clinical aspects, and can be queried directly through natural-language prompts.

We propose SLAP (Speaker contrastive Language-Audio Pretraining), the first framework designed for zero-shot inference of speaker and health attributes. SLAP converts heterogeneous annotations into natural language prompts and aligns them with audio through a contrastive objective, enabling out-of-distribution (OOD) generalization to unseen languages and clinical populations. Our contributions are threefold: \textbf{(i)} we designed a CLAP-style audio-language pretraining tailored for speaker, voice and health tasks, \textbf{(ii)} we conducted a large comprehensive evaluation of its kind, 38 tasks across 14 corpora in 7 languages, and \textbf{(iii)} we demonstrated, for the first time, zero-shot and out-of-domain generalization, especially for health-related speech analysis compared to speech foundation models, in addition to achieving state-of-the-art zero-shot and competitive linear probing results.
%we demonstrated better performances in linear probing in general that generalist speech foundation model and, for the first time, zero-shot and OOD generalization for health-related speech analysis.

\vspace{-0.5em}
\section{Methods}
\label{sec:Methods}
\vspace{-0.5em}
\subsection{Speaker Contrastive Language-Audio Pretraining}

\begin{figure}[t]
    \centering
    \includegraphics[width=\linewidth]{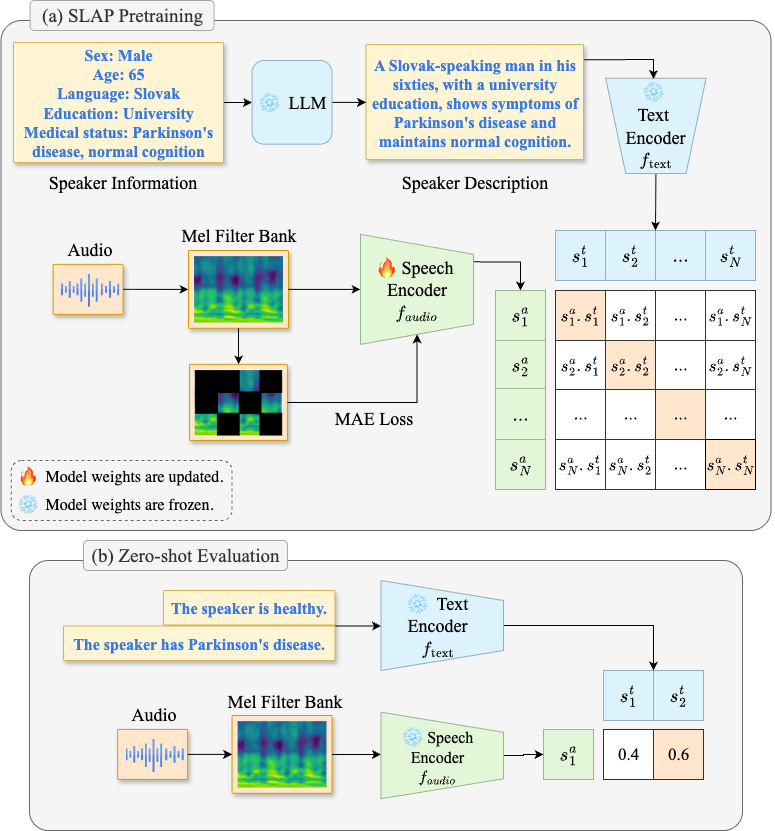}
    \caption{\textbf{SLAP pipeline.} \textbf{(a) Pretraining.} For each audio, an LLM generates a speaker description, and the Speech Encoder is trained with contrastive and self-supervised objectives. \textbf{(b) Zero-shot Evaluation}. A pair of text prompts are given to the Text Encoder. The class with the higher cosine similarity to the audio embedding is chosen.}
    \vspace{-1.5em}
    \label{fig:slap_archi}
\end{figure}

In Figure \ref{fig:slap_archi}, we illustrated the general architecture of our proposed speaker contrastive language-audio model. The input audio $X^{a}_{i}$ is converted into a Mel filter bank representation before being processed by the Speech Encoder. The speaker description $X^{t}_{i}$ is fed into the Text Encoder. The final audio and text embeddings $s_a^{(i)}$ and $s_t^{(i)}$ are obtained by passing the encoded features through learnable projection heads that map them to a shared $d$-dimensional embedding space:
\begin{equation}
s_a^{(i)} = \text{proj}_a\left(f_{\text{audio}}\left(X_i^a\right)\right), \quad s_t^{(i)} = \text{proj}_t\left(f_{\text{text}}\left(X_i^t\right)\right)
\end{equation}
where $f_{\text{audio}}$ and $f_{\text{text}}$ are the audio and text encoders. The similarity matrix $S$ is computed with cosine similarity:

\begin{equation}
S_{mn} = \frac{\langle s_a^{(m)}, s_t^{(n)} \rangle}{\|s_a^{(m)}\|_2 \cdot \|s_t^{(n)}\|_2}.
\end{equation}

Following the contrastive learning approach of CLAP \cite{elizalde2023clap}, the model was trained to maximize the similarity between matched audio-text pairs while minimizing similarity for unmatched pairs:

\begin{equation}
\begin{aligned}
\mathcal{L}_{\text{CLIP}} = -\frac{1}{2B} \sum_{i=1}^B \left[\log \frac{\exp(S_{ii} / \tau)}{\sum_{j=1}^B \exp(S_{ji} / \tau)} \right. \\
\left. + \log \frac{\exp(S_{ii} / \tau)}{\sum_{j=1}^B \exp(S_{ij} / \tau)}\right],
\end{aligned}
\end{equation}
where $\tau$ is a learnable temperature parameter for scaling the loss, $B$ is the batch size, and the embeddings are normalized to unit length before computing similarities. The two terms ensure bidirectional audio-text alignment in the shared embedding space.
Besides, we incorporated a masked autoencoder objective \cite{huang2022masked} for auxiliary self-supervised learning. Our total loss combines both objectives:
$\mathcal{L}_{\text{total}} = \lambda \mathcal{L}_{\text{CLAP}} + \mathcal{L}_{\text{MAE}}$
% ,
% \end{equation}
where $\lambda$ balances the contrastive and reconstruction objectives.

In this study, our audio backbone follows the ViT-B Masked Autoencoder \cite{huang2022masked} model initialized with weights pretrained on Audioset.
To generate speaker descriptions, we used an LLM to generate 3 alternative descriptions per individual by sampling from semantically equivalent terms. For instance, two alternatives for the same speaker from EWA-DB \cite{rusko2023ewa}: \textit{"A Slovak-speaking man in his sixties, with a university education, shows symptoms of Parkinson''s disease and maintains normal cognition."} and
\textit{"This mature adult male from Slovakia exhibits Parkinson’s disease while remaining cognitively healthy, and has completed university."}.

\vspace{-1.em}
\subsection{Downstream Tasks at Inference}
\label{sec:downstream_tasks}

We evaluated SLAP on a broad spectrum of speaker-centric downstream tasks, organized into three categories: demographics, voice characteristics, and health. All tasks are framed as binary classification. 

Demographic tasks (11 in total) target attributes such as age group and sex. Voice characteristics tasks (9) involve perceptual judgments such as strength, nasality, and standardized rating scales (GRBAS, CAPE-V). Health-related tasks (18) address clinical screening and diagnosis, including mental health (depression, anxiety, insomnia, fatigue, suicidal ideation) and neurological conditions (Parkinson’s disease, Alzheimer’s disease, ALS). In total, our evaluation covers 38 tasks across 14 datasets and 7 languages, providing a comprehensive benchmark for both in-domain, OOD performances and zero-shot generalization.

For \textit{supervised evaluation}, we extracted fixed-size representations with each approach and trained a Logistic Regression classifier on the training splits of each dataset. 
For \textit{zero-shot evaluation}, we followed the CLAP paradigm by comparing audio embeddings with text embeddings of task-specific prompts. For each binary classification task, we constructed positive and negative prompt pairs. Initial experiments revealed that negation (e.g., ``not depressed'') led to suboptimal performance at inference time. Instead, we used antonym pairs: \textit{"The speaker is depressed."} vs. \textit{"The speaker is healthy."}. Classification is performed by computing cosine similarities between the audio embeddings and the text embeddings, assigning the label with the higher similarity.

For both approaches, given that the audios have variable-length, we employed a sliding window approach: for utterances exceeding 10s, we extracted 10s windows with 5s overlaps, computed embeddings for each window, and aggregated the predictions via mean pooling.

\vspace{-1.2em}
\section{Experiments}
\vspace{-.5em}

\label{sec:Experiments}

\subsection{Pretraining Datasets and Speaker Descriptions}

We trained SLAP using the Adam optimizer with $\beta_1=0.9$ and $\beta_2=0.999$ on 4 NVIDIA  40GB A100 GPUs for 50k steps. We employed an exponential learning rate schedule with 2500 warmup steps, starting from $1 \times 10^{-4}$ with a learning rate decay factor of 0.99 applied after every 2000 step. The temperature parameter $\tau$ was initialized at $0.07$. % and learned during training as in original CLAP paper.
All audio are resampled to 16 kHz and transformed into log-mel spectrograms with 128 filterbanks, using a 25ms window size and 10ms hop size. There was no data augmentation. 

\begin{table}[t]
    \centering
    % \vspace{-6pt} 
    \caption{\textbf{Dataset statistics and metadata availability.} Checkmarks indicate available annotation types. Dem.: Demographics (age, sex), Voc.: Voice quality ratings, Med.: Medical/clinical assessments. (CN: Chinese, EN: English, FR: French, IT: Italian, SK: Slovak, SP: Spanish) \\ }
     \setlength{\tabcolsep}{1.6pt} 
    \label{tab:pretraining_table}
\begin{tabular}{lccccc}
\hline
  &  \#Spk  & Dur.(h) &  \multicolumn{3}{c}{Dimension} \\
 Dataset  &  &   & Dem. & Voc. & Med. \\
\hline
 \multicolumn{6}{l}{Pretraining} \\
\hline
Voxceleb1 \cite{nagrani2020voxceleb}   & 1251   & 351  & \checkmark &  & \\
Voxceleb2 \cite{chung18b_interspeech}   & 6112  & 2440   & \checkmark &  & \\
VOCES \cite{de2022virtual} (SP)   &  90  & 12  & \checkmark &  & \checkmark \\
CN-Celeb1 \cite{fan2020cn} (CN)  &  997  & 273 &  \checkmark &  & \\
PopGen \cite{riad2024automated} (FR) & 1667    &  135 & \checkmark &  & \checkmark \\
CLAC \cite{ramani2021clac} (EN)  & 1832 & 105   & \checkmark &  &  \\
EWA-DB \cite{rusko2023ewa} (SK) & 997   & 92  & \checkmark &  & \checkmark \\
Androids \cite{tao23_interspeech} (IT)  & 110   & 7.5  &  &  & \checkmark \\
\hline 
Total  & 13056   & 3415.5  \\
\hline
 \multicolumn{6}{l}{OOD test datasets} \\
 % \multicolumn{6}{l}{} \\
\hline
DreamVoiceDB \cite{hai24_interspeech} (EN)  & 899   & 221.5  &  & \checkmark & \\
PVQD \cite{walden2022perceptual} (EN)  & 296   & 2  & \checkmark & \checkmark & \checkmark \\
Lanzhou \cite{cai2022multi} (CN)  & 52   & 7  & \checkmark &  & \checkmark \\
SAP \cite{hasegawa-johnson24_sap} (EN)  &  424 & 304 & \checkmark & \checkmark \\
Neurovoz \cite{mendes2024neurovoz} (SP)  & 105   & 3.5  &  &  & \checkmark \\
Naples Voiced \cite{cesari2018new} (IT) & 208 & 0.3  & \checkmark & \checkmark & \checkmark \\
VOC-ALS \cite{dubbioso2024voice} (IT) & 153 & 5  &  &  & \checkmark \\
\hline 
Total  & 2137   & 543.3  \\
\hline
\end{tabular}
\vspace{-1.5em} 
\end{table}

We constructed our pretraining dataset from 9 diverse speech corpora, comprising over 3.4k hours of audio from more than 10k speakers, along with corresponding metadata and descriptions. The dataset comprises audio-text pairs extracted from Androids \cite{tao23_interspeech}, VOCES \cite{de2022virtual}, EWA-DB \cite{rusko2023ewa}, SAP \cite{hasegawa-johnson24_sap}, PopGen \cite{riad2024automated}, CLAC \cite{ramani2021clac}, VoxCeleb1 \cite{nagrani2020voxceleb}, VoxCeleb2 \cite{chung18b_interspeech}, and CN-Celeb1 \cite{fan2020cn}. These corpora provide rich speaker-level annotations including demographic information (age, sex), linguistic characteristics (e.g.: language, accent), health status (e.g.: Parkinson's, Alzheimer's, depression), and voice attributes (e.g.: roughness, intensity) that serve as textual supervision signals for our contrastive learning framework. Detailed statistics for each dataset, including the number of speakers, and available metadata fields, are provided in Table \ref{tab:pretraining_table}. 
To address data imbalance during pretraining, we resampled subsets of large datasets (VoxCeleb1, VoxCeleb2, CN-Celeb1, and CLAC) every 2000 steps.
Based on preliminary experiments, we generated descriptions with Gemma-3-27b-it \cite{team2025gemma} and we used Qwen3-Embedding-8B \cite{yang2025qwen3} to embed speaker textual descriptions to match. 

\vspace{-1.em}
\subsection{Downstream Tasks and Datasets}
\vspace{-0.2em}
We instantiated these 3 task categories using 14 corpora covering diverse populations and languages. We evaluated SLAP in both in-domain and OOD (out-of-distribution) conditions, reporting F1 scores on held-out test sets for zero-shot and supervised baselines.

\textbf{Demographics.} Sex and age classification tasks are evaluated across Naples Voiced \cite{cesari2018new}, VOCES \cite{de2022virtual}, PopGen \cite{riad2024automated}, PVQD \cite{walden2022perceptual}, and EWA-DB \cite{rusko2023ewa}.  

\textbf{Voice characteristics.} Perceptual assessments of vocal strength, nasality, and GRBAS ratings are drawn from DreamVoiceDB \cite{hai24_interspeech}, PVQD \cite{walden2022perceptual}, and SAP \cite{hasegawa-johnson24_sap}.  

\textbf{Health.} Tasks include mental health screening of depression, anxiety, insomnia, and fatigue in French general population (PopGen \cite{riad2024automated}), depression detection in Italian (Androids \cite{tao23_interspeech}) and Chinese (Lanzhou \cite{cai2022multi}), depression and suicidality detection in Spanish psychiatric patients (VOCES \cite{de2022virtual}), and neurological disease detection: Parkinson’s and Alzheimer’s in Slovak (EWA-DB \cite{rusko2023ewa}), Parkinson’s in Spanish (Neurovoz \cite{mendes2024neurovoz}) and ALS in Italian (VOC-ALS \cite{dubbioso2024voice}).

% 1. Types of domain (speaker, voice, health)
% 2. Metrics
% 3. Baseline comparisons: dummy, whisper, opensmile, AudioMAE
\vspace{-1.5em}
\section{Results and discussions}
\label{sec:Results}
\vspace{-1.em}
We compared SLAP with several baselines representing different paradigms in speaker tasks: OpenSMILE with eGeMAPS \cite{eyben2015geneva}, AudioMAE \cite{he2022masked} and the generalist foundation models WhisperM and WhisperL \cite{radford2023robust}.
% Note that we chose WhisperM over WhisperS and WhisperL due to its superior performance on our evaluation protocol.
We also pretrained our AudioMAE model, called SLAP-MAE, on the same pretraining datasets as SLAP to compare to pure self-supervised learning ($\lambda=0$). Both SLAP and SLAP-MAE was initialized with the pretrained weights of AudioMAE. For zero-shot approaches, we compared SLAP to two open-source CLAP models from \cite{wu2023large} \footnote{\url{https://huggingface.co/laion/clap-htsat-fused}}, the LAION CLAP-Fused model and an optimized version for music and speech referred as CLAP-MS \footnote{\url{https://huggingface.co/laion/larger_clap_music_and_speech}}. All supervised baselines were trained via linear probing on the train set of each dataset and evaluated on the test split, whereas zero-shot baselines are evaluated directly on the test sets with natural language prompts. 

% TODO complete. All supervised baselines are trained on available training splits, while zero-shot methods operate directly on test sets using natural language prompts.

\begin{table}[t]
    \centering
    \caption{\textbf{Classification results} comparing supervised and zero-shot approaches across different speaker attribute prediction tasks. All results are reported on the test sets with F1 scores. Best results are \textbf{bold} and second best results are \underline{underlined}. The Naive Classifier predicts the majority class. \\}
    \setlength{\tabcolsep}{1.65pt} 
    \label{tab:results}
    \begin{tabular}{lcccc}
    \hline
        \textbf{Dimension} & \textbf{Demog.} & \textbf{Voc.} & \textbf{Health} & \textbf{Total} \\
        \hline
        Naive Classifier  & 50.8 & 22.3 & 36.9 & 34.3 \\
        \hline
        % \multicolumn{5}{l}{In-domain evaluation} \\
        % \hline
        % \multicolumn{5}{l}{\quad \underline{Supervised Baselines}} \\
        % OpenSMILE  & xx & xx & xx & xx \\
        % Whisper  & xx & xx & xx & xx \\
        % Pretrained AudioMAE  & xx & xx & xx & xx \\
        % Finetuned AudioMAE  & xx & xx & xx & xx \\
        % \multicolumn{5}{l}{\quad \underline{Zero-shot Approaches}}  \\
        % Pure CLAP & xx & xx & xx & xx \\
        % SLAP (Ours) $\lambda=0$& \textbf{xx} & \textbf{xx} & \textbf{xx} & \textbf{xx} \\
        % SLAP (Ours) $\lambda=1$& \textbf{xx} & \textbf{xx} & \textbf{xx} & \textbf{xx} \\
        % \hline
        \multicolumn{5}{l}{Overall (In domain + OOD evaluation)} \\
        \hline
        \multicolumn{5}{l}{\quad \underline{Supervised Baselines}} \\
        OpenSMILE \cite{eyben2015geneva}  & 87.8 & 60.3 & 39.7 & 58.3 \\
        WhisperM \cite{radford2023robust} & \textbf{95.1} & \textbf{67.7} & \underline{56.1} & \textbf{70.1} \\
        WhisperL \cite{radford2023robust} & \underline{92.8} & 62.4 & 55.5 & 67.9 \\   
        
        AudioMAE \cite{he2022masked}  & 88.7 & 64.6 & 51.5 & 64.6 \\
        SLAP-MAE (ours, $\lambda=0$)  & 91.8 & \underline{66.7} & 51.8 & \underline{69.9} \\
        
        CLAP-Fused \cite{wu2023large} & 84.3 & 54.5 & 32.2  & 52.5 \\
        CLAP-MS \cite{wu2023large} & 76.4 & 60.8 & 31.7 &  52.5\\
        
        SLAP (ours, $\lambda=1$) & 91.4 & 65.2 & \textbf{57.9} & 69.3 \\

        \multicolumn{5}{l}{\quad \underline{Zero-shot Approaches}}  \\
        CLAP-Fused \cite{wu2023large} & 38.4 & 30.4 & 29.5 & 32.3 \\
        CLAP-MS \cite{wu2023large} & \underline{52.3} & \underline{35.0} & \underline{40.0} & \underline{42.4} \\
        SLAP (ours, $\lambda=1$) & \textbf{90.0} & \textbf{53.3 }& \textbf{51.2} & \textbf{62.9} \\

        \hline
        \multicolumn{5}{l}{OOD evaluation} \\
        \hline
        \multicolumn{5}{l}{\quad \underline{Supervised Baselines}} \\
        OpenSMILE \cite{eyben2015geneva} & 91.4 & 60.3 & 53.8 & 64.8 \\
        WhisperM \cite{radford2023robust} & \textbf{96.4} & \textbf{67.7} & 75.6 & \underline{76.2} \\
        WhisperL \cite{radford2023robust} & 89.9 & 62.4 & 77.1 & 72.9 \\   
        AudioMAE \cite{he2022masked}  & 89.6 & 64.6 & 71.3 & 72.0 \\
        SLAP-MAE (ours, $\lambda=0$) & \underline{91.6} & \underline{66.7} & \textbf{82.7} & \textbf{77.0} \\
        CLAP-Fused \cite{wu2023large} & 82.9 & 54.5 & 42.0 & 56.5 \\
        CLAP-MS \cite{wu2023large} & 77.0 & 60.8 & 41.8 & 58.2 \\
        SLAP (ours, $\lambda=1$) & 90.7 & 65.2 & \underline{79.0} & 74.9 \\

        \multicolumn{5}{l}{\quad \underline{Zero-shot Approaches}} \\
        CLAP-Fused \cite{wu2023large} & 33.3 & 30.4 & 23.2 & 28.8 \\
        CLAP-MS \cite{wu2023large} & \underline{52.8} & \underline{35.0} & \underline{47.1} & \underline{42.5}\\

        % SLAP (Ours) $\lambda=0$& \textbf{xx} & \textbf{xx} & \textbf{xx} & \textbf{xx} \\
        SLAP (ours, $\lambda=1$) & \textbf{89.6} & \textbf{53.3} & \textbf{58.5} & \textbf{62.6} \\

        \hline
         \
    \end{tabular}
    \vspace{-2em}
    \label{tab:results}
\end{table}

\begin{figure}[t]
    \centering
    \includegraphics[width=\linewidth]{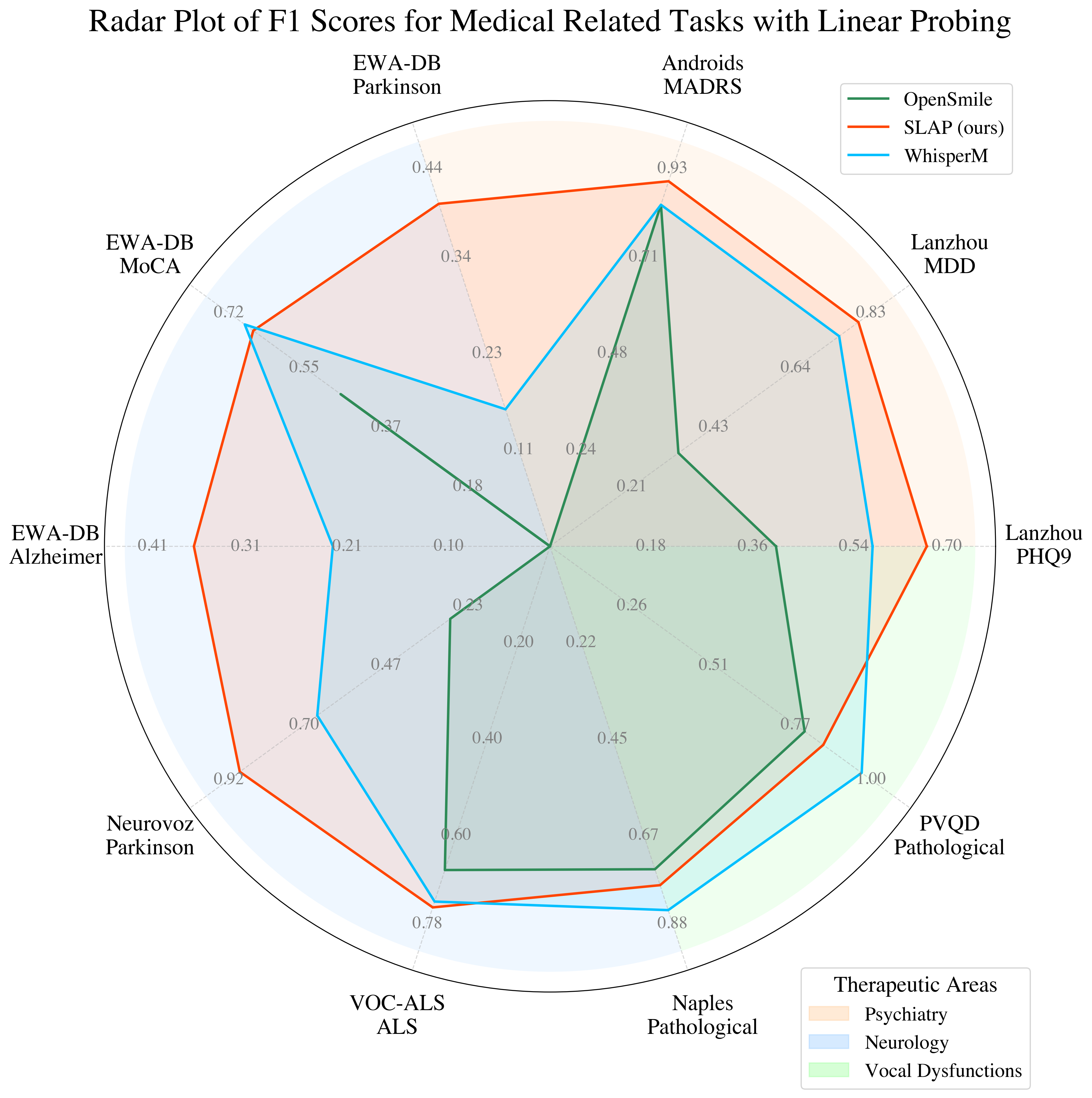}
    \caption{\textbf{Performances on open source medical tasks.} 
    \\F1 scores on psychiatry, neurology, and vocal dysfunctions.}
    \label{fig:radar}
    \vspace{-1.5em}
\end{figure}

Table \ref{tab:results} reports the classification results for supervised and zero-shot approaches, considering both OOD and overall (in- and OOD) performance. The results are averaged separately for each task category (demographics, vocal and health) as well as for all tasks. SLAP set state-of-the-art zero-shot results (90.0 F1 on demographics, 51.2 on health), approaching supervised performance. As mentioned in the introduction, the original CLAP model computes mel filter banks with hop sizes of 320ms and window sizes of 1024ms, which is not optimal to capture the fine-grained temporal structure essential for speech. In linear probing, WhisperM obtained the best overall results (F1 70.1), except on medical tasks where SLAP performed best (F1 57.9). Interestingly, SLAP-MAE surpassed SLAP in linear probing, highlighting the robustness of self-supervised pretraining. However, it cannot be evaluated in zero-shot setting. Combining contrastive and self-supervised pretraining enables SLAP to deliver state-of-the-art zero-shot results while remaining strong in linear probing. 

In Figure \ref{fig:radar}, we compared performances of SLAP, OpenSmile and WhisperM on all psychiatry, neurology, and vocal dysfunction tasks of open-source clinical datasets. SLAP outperformed WhisperM for the majority of tasks, and while WhisperM showed great performance on depression and vocal tasks overall, performances dropped on neurology tasks.

\vspace{-7pt}
\section{Conclusion}
\label{sec:Conclusion}
\vspace{-3pt}

SLAP demonstrated that combining contrastive language-audio pretraining with self-supervised learning enables both strong zero-shot performance and competitive supervised results on speaker and health-related tasks. By learning from heterogeneous speaker metadata with natural language supervision, the model achieved generalization across languages and clinical populations without task-specific training.
This capability addresses a critical challenge in clinical deployment: the need for models that can adapt to diverse populations and novel health conditions where labeled data is scarce or unavailable.

\newpage
\vfill\pagebreak

\section{Acknowledgments}
\label{sec:Acknowledgments}

This project was provided with computing AI and storage resources by GENCI at IDRIS thanks to the grant 20XX-AD010315777 on the supercomputer Jean Zay's V100/A100 partition. The authors are thankful to all the participants who volunteered for this research study. Without their active involvement, this study would not have been possible. The authors also would like to thank each of the speech pathology interns who helped with the participant recruitment and made sure that the protocol was completed successfully.

\small
\bibliographystyle{IEEEbib}
\bibliography{strings,refs}

\end{document}